\documentstyle[times,floats,aps,prb]{revtex}

\begin{document}

\draft
\twocolumn[\hsize\textwidth\columnwidth\hsize\csname
@twocolumnfalse\endcsname

\title {\bf Virtual-crystal approximation that works:  Locating a composition phase boundary in Pb(Zr$_{1-x}$Ti$_x$)O$_3$}
\author{Nicholas J. Ramer\cite{NJR} and Andrew M. Rappe}
\address{Department of Chemistry and Laboratory for Research on the
Structure of Matter\\ University of Pennsylvania, Philadelphia, PA
19104} 
\date{\today} 
\maketitle

\begin{abstract} 
We present a new method for modeling disordered solid solutions, based
on the virtual crystal approximation (VCA).  The VCA is a tractable
way of studying configurationally disordered systems; traditionally,
the potentials which represent atoms of two or more elements are
averaged into a composite atomic potential.  We have overcome
significant shortcomings of the standard VCA by developing a potential
which yields averaged atomic properties.  We perform the VCA on a ferroelectric
oxide, determining the energy differences between the high-temperature
rhombohedral, low-temperature rhombohedral and tetragonal phases of
Pb(Zr$_{1-x}$Ti$_{x}$)O$_3$ at $x=0.5$ and comparing these results to
superlattice calculations and experiment.  We then use our new method
to determine the preferred structural phase at $x=0.4$.  We find that
the low-temperature rhombohedral phase becomes the ground state at
$x=0.4$, in agreement with experimental findings.

\end{abstract} 
]

Throughout condensed matter science, solid solutions provide unique
properties which are inaccessible in pure materials.  Semiconductors
and ferroelectric materials are two important areas in which small
changes in atomic composition dramatically change material properties.
Theoretical examinations have been hampered by the difficulty of
incorporating inhomogeneities into perfectly periodic systems.
Several theoretical approaches have been formulated to address this
specific issue, including the virtual crystal approximation
(VCA)~\cite{VCA1,VCA2} and coherent potential approximation
(CPA).~\cite{Soven1,Soven2,CPA4}

The VCA has been performed for over 30 years with mixed results; a
composite potential is constructed which represents the average of the
component atoms comprising the inhomogeneity.  Previous
implementations of the VCA have studied how structural and electronic
properties of semiconducting and ferromagnetic alloys depend on
composition.  These studies typically provide qualitative and in some
cases quantitative agreement with experimental and large-scale
solid-state calculations.~\cite{Papa,Slavenburg} In some cases
however, these methods have given unphysical
results.~\cite{Chen,Bellaiche} The inability of these traditional
versions of the VCA to represent differences in chemical bonding and
ionicity accurately has been suggested as the reason for these
erroneous results.~\cite{Chen} In principle, the VCA affords great
versatility, applicable wherever compositional inhomogeneity can
treated in an averaged way.  On its own, the VCA does not describe the
different local atomic environment in inhomogeneous materials.  Local
features can be studied by performing a VCA calculation and adding in
the differences between true atomic potentials and the VCA using a
linear response method.~\cite{Stefano,Marzari}

For predictions of alloy densities of states, the CPA has achieved
considerable success.~\cite{CPA1,CPA2,CPA3} In this approach, a
Green's function is constructed such that an electron encountering
either element of the alloy will not scatter on average.  Recently,
nearest-neighbor charge correlations have been included into CPA
calculations.~\cite{DDJohnson} Despite this success, it is not
convenient to perform first-principles calculations within the CPA
which include self-consistency, Hellmann-Feynman forces and atomic
motion.  Therefore, considerable effort has been exerted toward the
development of a VCA appropriate to first-principles calculations.

In response to the shortcomings of traditional approaches to the VCA,
we have formulated a new approach that preserves important atomic
properties (atomic orbital energies, ionicity, polarizability, and
average size of the virtual atom).  The VCA method presented in this
paper represents a considerable departure from other first-principles
VCA treatments.  A variety of calculations have been performed by
averaging pseudopotentials of the component atoms and finding
self-consistent wave functions and charge density in the presence of
this average potential.  As we discuss below, the nonlinearity of the
self-consistent Kohn-Sham equation insures that the average potential
does not yield averaged properties.  Accordingly, we construct a
potential which does give correctly averaged atomic eigenvalues for a
wide variety of electronic configurations.  We anticipate that our new
VCA method, because it provides correctly averaged atomic {\em
properties}, will yield more accurate virtual atoms and provide
greater agreement with superlattice solid-state calculations than
traditional VCAs based on averaged {\em potentials}.  The VCA
construction methods are described for the combination of two atoms
according to A$_{1-x}$B$_{x}$.  In addition, we restrict our
description to homovalent atoms.  These methods can be generalized to
the averaging of more than two atoms and heterovalency.  It is
important to emphasize that since we construct a single potential at
the atomic level, the use of our VCA pseudopotentials in solid-state
calculations requires no additional computational effort.

We have chosen to assess the accuracy of the two methods to reproduce
superlattice calculations by computing the relative energy differences
between different phases of a ferroelectric solid solution ($x=0.5$)
of PbZrO$_3$ and PbTiO$_3$.  This composition lies very near the
tetragonal-rhombohedral morphotropic phase boundary of this material.
After establishing the accuracy of our new method, we use it to
predict the energetic ordering of Pb(Zr$_{1-x}$Ti$_x$)O$_3$ (PZT)
phases at another value of $x$.  We have previously demonstrated the
effectiveness of our new approach for phase stability and for
strain-induced phase transitions.~\cite{Williamsburg99} In this paper,
we perform a more complete treatment of the PZT work, involving three
structural phases and studying a compositional phase transition.

{\it{ADVCA (Averaged Descreened Potentials VCA)}}: In this approach,
two independent atomic pseudopotentials ($\widehat{V}_{\mathrm
PS}^{\mathrm A}$ and $\widehat{V}_{\mathrm PS}^{\mathrm B}$) are
constructed.  These descreened potentials are then averaged according
to:

\begin{eqnarray}
\widehat{V}_{\mathrm PS}^{\mathrm ADVCA} = (1-x) \widehat{V}_
{\mathrm PS}^{\mathrm A} + x \widehat{V}_{\mathrm PS}^{\mathrm B}.
\end{eqnarray}

\noindent In order to express the above semilocal VCA potential in a
nonlocal form,~\cite{KB} we must also know the semilocal
pseudo-wave-functions.  Therefore, we solve for the reference-state
pseudo-wave-functions of $\widehat{V}^{\mathrm ADVCA}_{\mathrm PS}$
and use these to construct the nonlocal potential.

{\it{RRVCA (New Method)}}: The ADVCA construction does not guarantee
that the virtual atom will have atomic properties which are the
correct average of the component atoms.  Therefore, we impose averaged
properties as a new criterion in our RRVCA construction.  We define an
averaged eigenvalue for the $l$-th state of the virtual atom,

\begin{eqnarray}
\varepsilon^{\mathrm AVG}_{l} = (1-x)\varepsilon^{\mathrm A}_{nl} +
 x\varepsilon^{\mathrm B}_{n'l},
\end{eqnarray}

\noindent where $\varepsilon^{\mathrm A}_{nl}$ and
$\varepsilon^{\mathrm B}_{n'l}$ are the valence all-electron
eigenvalues for the $nl$- and $n'l$-th states of atoms A and B, and
$n$ need not equal $n'$.  We seek a VCA potential which guarantees
that $\varepsilon^{\mathrm RRVCA}_{l}=\varepsilon^{\mathrm AVG}_{l}$
for the reference state.  As a first step, the bare nuclear Coulombic
potentials of the component atoms are averaged:

\begin{eqnarray}
V_{\mathrm AE}^{\mathrm RRVCA}&=&\frac{-2((1-x) Z_{\mathrm
AE}^{\mathrm A} + x Z_{\mathrm AE}^{\mathrm B})}{r}.
\end{eqnarray}

In addition, we determine a core charge density that is the weighted
average of the core charge densities of the all-electron component
atoms ($\rho^{\mathrm A}_{\mathrm core}$ and $\rho^{\mathrm
B}_{\mathrm core}$),

\begin{eqnarray}
\rho^{\mathrm RRVCA}_{\mathrm core} = (1-x)\rho^{\mathrm A}_{\mathrm
core}+x\rho^{\mathrm B}_{\mathrm core}.
\end{eqnarray}

By averaging the core charge densities and all-electron potentials, we
insure that the resulting virtual atom will possess the proper
averaged size of the two component atoms.  Using this nuclear
potential and frozen-core charge density, we find new all-electron
wave functions for the virtual atom valence states.  We accomplish
this by completing a self-consistent inward solve~\cite{Froese} for
the valence wave functions according to

\begin{enumerate}
\item $\varepsilon^{\mathrm RRVCA}_{l}=\varepsilon^{\mathrm AVG}_{l}$,
\item $\phi^{\mathrm RRVCA}_{l}(r)\rightarrow 0$ as $r\rightarrow 
\infty$,
\item $\int_{r_c}^{\infty}\left|\phi^{\mathrm RRVCA}_{l}(r)\right|^2 
r^2 dr=\\ (1-x)\int_{r_c}^{\infty}\left|\phi^{\mathrm A}_{nl}(r)
\right|^2 r^2 dr+x\int_{r_c}^{\infty}\left|\phi^{\mathrm B}_{n'l}
(r)\right|^2 r^2 dr$,
\end{enumerate}

\noindent where $r_c$ is the pseudopotential core radius for the
valence state.  Operationally, we complete an inward solve well below
$r_c$, insuring accurate first and second derivative determination for
all values of $r$ greater than $r_c$.  These derivatives are required
for the optimized pseudopotential construction.~\cite{RappePS}

Since the form of these all-electron RRVCA wave functions will be
modified within the core region when the wave functions are
pseudized,~\cite{RappePS} the Kohn-Sham equations need not be solved
between $r$=0 and $r$=$r_c$.  We construct the remainder of the
wave function as a smooth analytic form insuring norm-conservation and
agreement with the inward solution.  A new valence charge density is
constructed, and the process is iterated to self-consistency.  These
wave functions are self-consistent solutions to the Kohn-Sham equation
outside $r_c$, with eigenvalues and total norm outside $r_c$ which are
exactly the average of the component atoms.  Optimized semilocal
pseudopotentials are then constructed for this set of wave functions
and eigenvalues.  It is important to note that any method for
pseudopotential construction may be used because the averaging occurs
at the all-electron level.

\begin{table}[t]
\caption{Construction parameters for the Zr, Ti and Zr$_{1-x}$Ti$_x$
virtual crystal (VCA) designed nonlocal pseudopotentials.  The Zr and
Ti potentials were generated with the methods described in references
\protect\onlinecite{RappePS} and \protect\onlinecite{RRPS}.  The
VCA potentials were generated using the methods described in text.
Core radii ($r_c$) are in atomic units, $q_c$ parameters are in
Ry$^{1/2}$, step widths are in atomic units and step heights are in
Ry.}

\begin{tabular}{lcdcd}
    &       &       &Step &Step   \\
Atom&$r_{c}$&$q_{c}$&Range&Height \\ \hline
Zr ($4s^{2}4p^{6}4d^{0})$&1.51,1.51,1.90&7.07&0.00--1.51 &
0.77   \\ 
    &             &        &     &       \\
Ti ($3s^{2}3p^{6}3d^{0}$)&1.32,1.20,1.40&7.07&0.00--1.15 &
5.49   \\
    &             &             &     &       \\
ADVCA ($s^{2}p^{6}d^{0}$)&\multicolumn{4}{c}{generated by averaging} \\ 
$x=0.5$&\multicolumn{4}{c}{Zr and Ti potentials}\\
    &             &       &         &       \\

RRVCA ($s^{2}p^{6}d^{0}$)&1.38,1.51,1.40&7.07&0.00--0.56 &
1.18   \\ 
$x=0.5$&                                 &              &0.56--0.79 &
1.34   \\ 
    &             &       &         &       \\
RRVCA ($s^{2}p^{6}d^{0}$)&1.27,1.38,1.61&7.07&0.00--1.25 &
10.32   \\
$x=0.4$&             &       &      &       \\
\end{tabular}
\end{table}

The pseudopotential construction procedure guarantees the agreement of
the RRVCA eigenvalues with the averaged all-electron eigenvalues for
the reference state {\em only}.  To improve the transferability of the
RRVCA potential to other electronic configurations, we have employed
the designed nonlocal pseudopotential method.~\cite{RRPS} This method
relies on the arbitrariness of the separation between the local and
semilocal correction terms that are used in the nonlocal potential.
By adjusting the separation, we can dramatically improve the
transferability of the RRVCA potential.  Improving the agreement with
the all-electron atomic orbital energies for a range of electronic
configurations insures correct ionicity and polarizability of the
virtual atom.  It is important to note that any method of
pseudopotential construction sufficiently flexible to achieve
eigenvalue agreement for many electronic configurations will also
improve the quality of the virtual atom.

\begin{table}[t]
\caption{Configuration testing for Zr, Ti and Zr$_{1-x}$Ti$_x$
virtual crystal (VCA) atoms generated with the ADVCA and RRVCA methods
described in text.  Averaged eigenvalues and total energy differences
are given for the Zr and Ti all-electron (AE) and component
pseudopotentials (PS).  Absolute errors are computed as a difference
from the averaged component pseudopotential results.  All energies are
in Ry.}

\begin{tabular}{crrrr}
     &$x=0.5$&$x=0.5$&$x=0.5$&$x=0.5$ \\
     &\multicolumn{1}{c}{AE}&\multicolumn{1}{c}{PS}&\multicolumn{1}{c}{ADVCA}
&\multicolumn{1}{c}{RRVCA}\\ 
State&\multicolumn{1}{c}{Energy}&\multicolumn{1}{c}{Energy}&
\multicolumn{1}{c}{Error}&\multicolumn{1}{c}{Error}\\ \hline  
$s^2$&-7.5701&-7.5701& 0.2108& 0.0000\\  
$p^6$&-5.9530&-5.9530& 0.1554& 0.0000\\  
$s^0$&-2.5554&-2.5581& 0.0437&-0.0038\\  
$d^0$&-3.4488&-3.4488& 0.2607& 0.0000\\
$\Delta E_{\rm{tot}}$& 0.0000& 0.0000& 0.0000& 0.0000\\
 & & & & \\
$s^2$&-6.7832&-6.7808& 0.1942& 0.0016\\  
$p^6$&-5.1737&-5.1720& 0.1398& 0.0015\\  
$s^1$&-2.0292&-2.0304& 0.0335&-0.0009\\  
$d^0$&-2.7037&-2.7034& 0.2362& 0.0041\\
$\Delta E_{\rm{tot}}$&-2.3049&-2.3067&0.0387&-0.0019\\
 & & & & \\
$s^2$&-6.3611&-6.3652& 0.1434&-0.0031\\  
$p^6$&-4.7701&-4.7744& 0.0923&-0.0049\\  
$s^0$&-1.8390&-1.8399& 0.0147& 0.0000\\  
$d^1$&-2.3479&-2.3571& 0.1782&-0.0033\\
$\Delta E_{\rm{tot}}$&-2.8876&-2.8936& 0.2201&-0.0012\\
 & & & & \\
$s^2$&-5.6569&-5.6586& 0.1266&-0.0029\\  
$p^6$&-4.0711&-4.0734& 0.0764&-0.0047\\  
$s^1$&-1.3449&-1.3453& 0.0119& 0.0011\\  
$d^1$&-1.6814&-1.6889& 0.1549&-0.0018\\
$\Delta E_{\rm{tot}}$&-4.4856&-4.4923& 0.2348&-0.0004\\
 & & & & \\
$s^2$&-4.1762&-4.1730& 0.0398&-0.0077\\  
$p^6$&-2.6089&-2.6064&-0.0059&-0.0106\\  
$s^2$&-0.3301&-0.3293&-0.0097& 0.0018\\  
$d^2$&-0.3205&-0.3239& 0.0586&-0.0072\\ 
$\Delta E_{\rm{tot}}$&-6.2574&-6.2696& 0.3397&-0.0037\\
\end{tabular}
\end{table}

We have applied the ADVCA and RRVCA methods to the Zr$_{1-x}$Ti$_x$
virtual atom at $x=0.5$.  All atomic energy calculations were done
within the local density approximation, and the
optimized~\cite{RappePS} and designed nonlocal
pseudopotential~\cite{RRPS} methods were used.  The generation
parameters for the component potentials and the VCA potentials are
included in Table I.  For all atoms, semi-core states were included as
valence.  It is important to note that although we have included
multiple $s$-channel states, only one $s$ nonlocal projector is
included.  For all atoms we have used the $s$-potential as the local
potential onto which we add one or two square-step augmentation
operators.  The transferability data for the ADVCA and RRVCA atoms are
presented in Table II.  For comparison we have also included the
all-electron averaged eigenvalues and the averaged component
pseudopotential eigenvalues.  All errors in the VCA methods are
computed as the difference from the averaged component pseudopotential
eigenvalues.  From the table it is clear that the RRVCA provides the
more transferable potentials; the RRVCA gives atomic properties which
are orders of magnitude closer to the weighted average of the
component atoms than the ADVCA.  We now test both VCA potentials,
using them in solid-state calculations and comparing to superlattice
results.  We have completed first-principles calculations using
density functional theory and the local density approximation on a
Pb(Zr$_{0.5}$Ti$_{0.5}$)O$_3$ superlattice with stacking in the (111)
direction~\cite{Ferro98} and (011) direction. The electronic wave
functions are expanded in a plane-wave basis using a cutoff energy of
50 Ry.  In addition to the semi-core states mentioned above, the 5$d$
shell is included in the Pb potential.  Scalar relativistic effects
are included in the generation of the Pb pseudopotential.  Brillouin
zone integrations are approximated accurately as sums on a
4$\times$4$\times$4 Monkhorst-Pack $k$-point mesh.~\cite{MNKPCK}

The experimentally determined phase diagram for PZT shows two phase
boundaries relevant to the present work.  A nearly
temperature-independent boundary at $x\approx 0.45$ separates the
Ti-rich tetragonal $P4mm$ phase from the Zr-rich rhombohedral ($R3c$
and $R3m$) phases.~\cite{Jaffe,PZTJaffe} Around 400-500K, the
rhombohedral region exhibits a boundary between $R3c$
(low-temperature) and $R3m$ (high-temperature) phases, which depends
weakly on composition.~\cite{Michel,Glazer} The $R3c$ phase shows
complex oxygen octahedral tilting, which doubles the primitive unit
cell to ten atoms.~\cite{Clarke}

Table III contains the $x=0.5$ PZT superlattice equilibrium cell
dimensions and energy differences for the tetragonal, low- and
high-temperature rhombohedral phases for both cation orderings. For
all rhombohedral calculations, we have neglected the small shear
relaxations.  Energy differences have been computed relative to the
tetragonal phase.  For the (111) ordering, we find that the low- and
high-temperature rhombohedral phases are 0.035~eV and 0.05~eV higher
in energy than the tetragonal phase.  For the (011) ordering, we find
the low- and high-temperature rhombohedral phases are 0.093~eV and
0.115~eV above the tetragonal ground state.  For both orderings, the
low-temperature rhombohedral phase is lower than the high-temperature
phase.

For both VCA methods, we have completed full electronic and structural
relaxations for five-atom unit cells which possess tetragonal and
$R3m$ symmetries.  For the low-temperature rhombohedral phase, we have
studied a 10-atom unit cell which possesses $R3c$ symmetry.  In Table
III, we report the results of the ADVCA.  ADVCA predicts that the
ground-state phase for this composition of PZT is the high-temperature
rhombohedral phase, in direct opposition to the superlattice
calculations and experimental observations.  In addition we find that
ADVCA significantly underestimates the $c/a$ ratio for the tetragonal
phase.  We attribute the lack of agreement to the poor quality of the
ADVCA virtual atom.  The second to last column of Table III contains
the results from the RRVCA.  We find the same energy ordering as the
superlattice calculations and experiment, with the low- and
high-temperature phases 0.05~eV and 0.072~eV higher than the
tetragonal phase.  Finally, we find excellent agreement in the
structural parameters between the RRVCA and superlattice calculations.
We attribute the quantitative agreement in the energetic and
structural results to the improved quality of the RRVCA virtual atom.

\begin{table}[t]
\caption{Solid-state calculation results for superlattice and virtual
crystals of $x=0.5$ and $x=0.4$ compositions of
Pb(Zr$_{1-x}$Ti$_x$)O$_3$ with (111) and (011) orderings.  Energy
differences are relative to the tetragonal phase in all cases.
Energies and structural parameters are given for a 40-atom unit cell.}

\begin{tabular}{lcrrrrr}
     & &\multicolumn{2}{c}{Superlattice}& & \\ \cline{3-4}
 & &\multicolumn{1}{c}{$x=0.5$}&\multicolumn{1}{c}{$x=0.5$}&\multicolumn{1}{c}{$x=0.5$}&\multicolumn{1}{c}{$x=0.5$}&\multicolumn{1}{c}{$x=0.4$} \\
Structure& &\multicolumn{1}{c}{(111)}&\multicolumn{1}{c}{(011)}&\multicolumn{1}{c}{ADVCA}&\multicolumn{1}{c}{RRVCA}&\multicolumn{1}{c}{RRVCA} \\
\hline
Rhombohedral        &$a$       & 8.052& 8.029& 8.077& 8.043& 8.033\\  
    ($R3m$)         &$\Delta E$& 0.050& 0.115&-0.041& 0.072& 0.048\\
 & & & & & & \\
Tetragonal          &$c$       & 8.199& 8.219& 8.131& 8.210& 8.199\\
   ($P4mm$)         &$a$       & 7.869& 7.989& 8.054& 7.952& 7.957\\
	            &$c/a$     & 1.041& 1.029& 1.010& 1.033& 1.030\\
                    &$\Delta E$& 0.000& 0.000& 0.000& 0.000& 0.000\\
 & & & & & & \\
Rhombohedral        &$a$       & 8.026& 8.004& 8.046& 8.007& 8.019\\  
           ($R3c$)  &$\Delta E$& 0.035& 0.093&-0.067& 0.050&-0.044\\ 
\end{tabular}
\end{table}

Having demonstrated the effectiveness of the RRVCA and the inability
of the ADVCA to reproduce the superlattice results for $x=0.5$, we use
the RRVCA to examine further the compositional phase transition in
PZT.  We have chosen to study the $x=0.4$ composition of PZT, since
the ground-state structure for this composition lies in the
low-temperature rhombohedral phase region of the phase diagram.  We
have constructed a $x=0.4$ Zr$_{1-x}$Ti$_x$ potential using the RRVCA
according to the parameters in Table I.  The transferability of this
potential is equal to that of the $x=0.5$ RRVCA potential.  We have
completed structural relaxations of the two five-atom unit cell
structures described above as well as the ten-atom low-temperature
rhombohedral phase.  The last column of Table III contains the
structural parameters for the three phases and energy differences
relative to the tetragonal phase.  We find that moving from $x=0.5$ to
$x=0.4$ composition shifts both rhombohedral phases lower in energy
relative to the tetragonal phase.  The high-temperature rhombohedral
phase energy drops by approximately 0.012 eV and the low-temperature
phase decreases by 0.094 eV, becoming the ground-state structure for
the $x=0.4$ composition.  This result is in direct agreement with the
experimental findings and demonstrates the ability of the RRVCA to
predict compositional phase transitions.

In this paper we have presented a new method for constructing virtual
crystal pseudopotentials and have applied this method to the Ti and Zr
atoms.  Our new method is based on the self-consistent determination
of averaged all-electron properties which insures exact agreement of
the proper averaged all-electron eigenvalues for the reference
configuration of the virtual atom.  This procedure also yields the
proper averaged atomic size.  To improve the electronic properties of
the virtual atom at all other electronic configurations, we employed
the designed nonlocal pseudopotential approach.  We have compared our
new method to a more traditional one in which the descreened
pseudopotentials are averaged, computing the relative energy
differences for the high-temperature rhombohedral, low-temperature
rhombohedral and tetragonal phases of Pb(Zr$_{1-x}$Ti$_{x}$)O$_3$ and
comparing those results to superlattice calculations with two
different cationic orderings.  For this more traditional method, we
find neither quantitative nor qualitative agreement with superlattice
results.  However, for our new method, we find excellent agreement not
only in the energetics of the three phases but also the structural
parameters.  Finally we use our new method to predict the
compositional phase transition between $x=0.5$ and $x=0.4$
Pb(Zr$_{1-x}$Ti$_{x}$)O$_3$.  Our new method predicts that the
low-temperature rhombohedral phase is the ground-state structure for
the $x=0.4$ composition in agreement with experiment. This calculation
represents the first {\em ab initio} determination of a compositional
phase boundary in an ferroelectric oxide.

We have recently become aware of an alternative approach to improving
the first-principles VCA.  Bellaiche and
Vanderbilt~\cite{BellaichePRB} apply the nonlocal projectors of all
the component atoms at the site of the virtual atom.  Using ultrasoft
pseudopotentials,~\cite{USPSP} they find accurate piezoelectric and
ferroelectric properties for Pb(Zr$_{0.5}$Ti$_{0.5}$)O$_3$ using their
approach.

The authors would like to thank Ilya Grinberg for his help with
programming the virtual crystal methods.  In addition, we would like
thank David Pettifor, Anthony Paxton and Christian Elsasser for
valuable discussions.  This work was supported by NSF grant DMR
97-02514 and the Petroleum Research Fund of the American Chemical
Society (Grant No. 32007-G5) as well as the Laboratory for Research on
the Structure of Matter and the Research Foundation at the University
of Pennsylvania.  Computational support was provided by the San Diego
Supercomputer Center and the National Center for Supercomputing
Applications.


\begin{thebibliography}{99}

\bibitem[*]{NJR} Present address:  Department of Chemistry, Long Island University - C. W. Post Campus, Brookville, NY 11548.

\bibitem{VCA1} L. N{\o}rdheim, {\em Ann. Phys.\/} (Leipzig) {\bf 9}, 
607 (1931).

\bibitem{VCA2} T. Muto, {\em Sci. Pap. Inst. Phys. Chem. Res.} (Jpn.)
{\bf 34}, 377 (1938).

\bibitem{Soven1} P. Soven, {\em Phys. Rev.} {\bf 156}, 809 (1967).

\bibitem{Soven2} P. Soven, {\em Phys. Rev.} {\bf 178}, 1136 (1969).

\bibitem{CPA4} J. S. Faulkner, {\em Prog. Mat. Sci.} {\bf 27}, 187
(1982).

\bibitem{Papa} D. A. Papaconstantopoulos and W. E. Pickett, {\em Phys.
Rev. B\/} {\bf 57}, 12751 (1998).

\bibitem{Slavenburg} P. Slavenburg, {\em Phys. Rev. B\/} {\bf 55},
16110 (1997).

\bibitem{Chen} C. Chen, E. G. Wang, Y. M. Gu, D. M. Bylander and
L. Kleinman, {\em Phys. Rev. B} {\bf 57}, 3753 (1998).

\bibitem{Bellaiche} L. Bellaiche, S.-H. Wei and A. Zunger, {\em Appl. 
Phys. Lett.\/} {\bf 70}, 3558 (1997).

\bibitem{Stefano} S. de Gironcoli, P. Giannozzi and S. Baroni, {\em
Phys. Rev. Lett.} {\bf 66}, 2116 (1991).

\bibitem{Marzari} N. Marzari, S. de Gironcoli and S. Baroni, {\em
Phys. Rev. Lett.\/} {\bf 72}, 4001 (1994).

\bibitem{CPA1} G. M. Stocks, W. M. Temmerman and B. L. Gy{\"o}rffy, {\em
Phys. Rev. Lett.} {\bf 41}, 339 (1978).

\bibitem{CPA2} J. S. Faulkner and G. M. Stocks, {\em Phys. Rev. B}
{\bf 21}, 3222 (1980).

\bibitem{CPA3} J. S. Faulkner and G. M. Stocks, {\em Phys. Rev. B}
{\bf 23}, 5628 (1981).

\bibitem{DDJohnson} F. J. Pinski, J. B. Staunton and D. D. Johnson,
{\em Phys. Rev. B} {\bf 57}, 15177 (1998).

\bibitem{Williamsburg99} N. J. Ramer and A. M. Rappe, {\em
J. Phys. Chem. Solids}, {\bf 61}, 315 (2000).

\bibitem{KB} L. Kleinman and D. M. Bylander, {\em Phys. Rev. Lett.\/}
{\bf 48}, 1425 (1982).

\bibitem{Froese} C. Froese, {\em Can. J. Phys.\/} {\bf 41}, 1895
(1963).

\bibitem{RappePS} A. M. Rappe, K. M. Rabe, E. Kaxiras and
J. D. Joannopoulos, {\em Phys. Rev. B\/} {\bf 41}, 1227 (1990).

\bibitem{RRPS} N. J. Ramer and A. M. Rappe, {\em Phys. Rev. B\/} {\bf
59}, 12471 (1999).

\bibitem{Ferro98} N. J. Ramer, S. P. Lewis, E. J. Mele and
A. M. Rappe, in: R. E. Cohen (Ed.), {\em First-Principles Calculations
for Ferroelectrics - Fifth Williamsburg Workshop}. AIP, Woodbury, NY,
1998, p. 156.

\bibitem{MNKPCK} H. J. Monkhorst and J. D. Pack, {\em Phys. Rev. B\/}
{\bf 13}, 5188 (1976).

\bibitem{Jaffe} B. Jaffe, W. R. Cook Jr. and H. Jaffe, {\em 
Piezoelectric Ceramics\/}. TechBooks, Marietta, OH (1990). 

\bibitem{PZTJaffe} B. Jaffe, R. S. Roth and S. Marzullo, {\em
J. Res. Nat. Bur. Stand.} {\bf 55}, 239 (1955).

\bibitem{Michel} C. Michel, J.-M. Moreau, G. D. Achenbach, R. Gerson and W. J. James, {\em Sol. State Commun.} {\bf 7}, 865 (1969).

\bibitem{Glazer} A. M. Glazer, {\em Acta Cryst. B} {\bf 28}, 3384 (1972).

\bibitem{Clarke} R. Clarke and A. M. Glazer, {\em Ferroelectrics} {\bf 12}, 207 (1976).

\bibitem{BellaichePRB} L. Bellaiche and D. Vanderbilt, {\em
Phys. Rev. B}, {\bf 61}, 7877 (2000).

\bibitem{USPSP} D. Vanderbilt, {\em Phys. Rev. B} {\bf 41}, 7892
(1990).

\end{thebibliography}
\end{document}